\documentclass[]{elsart}
\usepackage{graphicx}
\usepackage{amssymb}
\usepackage{epstopdf}
\usepackage{array}
\usepackage{subfigure}
\usepackage{url}

\begin{document}
\begin{frontmatter}\
\title{Slime mould electronic oscillators}
\author{Andrew Adamatzky}
\address{University of the West of England, Bristol, BS 16 1QY, United Kingdom; Email:andrew.adamatzky@uwe.ac.uk}


\maketitle

\begin{abstract}

We construct electronic oscillator from acellular slime mould \emph{Physarum polycephalum}. The slime mould oscillator is made of two electrodes connected by a protoplasmic tube of the living slime mould. A protoplasmic tube has an average resistance of 3~MOhm. The tube's resistance is changing over time due to peristaltic contractile activity of
the tube. The resistance of the protoplasmic tube oscillates with average period of  73~sec and average amplitude of 
0.6~MOhm.   We present experimental laboratory results on dynamics of Physarum oscillator under direct current voltage up to 15~V and speculate that slime mould \emph{P. polycephalum} can be employed as a living electrical oscillator in biological and hybrid circuits. 

\end{abstract}

\end{frontmatter}

\section{Introduction}

\vspace{0.5cm}

\begin{quote}
``A device without an oscillator either doesn't do anything or expects to be driven by something else (which probably contains an oscillator).''\\
Horowitz and Hill, The Art of Electronics, 1980.
\end{quote}

\vspace{0.5cm}

The plasmodium of \emph{Physarum polycephalum} (Order \emph{Physarales}, class \emph{Myxomecetes}, subclass \emph{Myxogastromycetidae}) is a single cell, visible with the naked eye, with many diploid nuclei. The plasmodium feeds on bacteria and microscopic food particles by endocytosis. When placed in an environment with distributed sources of nutrients the plasmodium forms a network of protoplasmic tubes connecting the food sources. The topology of the plasmodium's protoplasmic network optimises the plasmodium's harvesting of nutrient resource from the scattered sources of nutrients and makes more efficient the transport of intra- cellular components~\cite{nakagaki_2000}. 
In \cite{adamatzky_physarummachines} we have shown how to construct specialised and general purpose massively-parallel amorphous computers from the plasmodium (slime mould) of \emph{P. polycephalum} that are capable of solving problems of computational geometry, graph-theory and logic. Plasmodium's foraging behaviour can be  interpreted as a computation~\cite{nakagaki_2000,nakagaki_2001a,nakagaki_iima_2007}:  data are represented by spatial of attractants and repellents, and  results are represented by structure of protoplasmic  
network~\cite{adamatzky_physarummachines}.  
Plasmodium can solve computational problems with natural parallelism, e.g. related to shortest 
path~\cite{nakagaki_2001a} and hierarchies of planar proximity graphs~\cite{adamatzky_ppl_2008}, computation of plane tessellations~\cite{shirakawa}, execution of logical computing schemes~\cite{tsuda2004,adamatzky_gates}, and natural implementation of spatial logic and process algebra~\cite{schumann_adamatzky_2009}.  

In the framework of our ``Physarum Chip'' EU project~\cite{adamatzky_phychip} we aim to experimentally implement a working prototype of a Physarum based general purpose computer. This computer will combine self-growing computing circuits made of a living slime mould with conventional electronic components. Data and control inputs to the Physarum Chip will be implemented via chemical, mechanical and optical means.   Aiming to develop a component base of future Physarum computers we designed Physarum tactile  sensor~\cite{adamatzky_tactilesensor} and
 undertook foundational studies towards fabrication of  slime mould chemical sensors  (Physarum nose)~\cite{delacycostello_2013, whitting_2013}, Physarum memristive devices~\cite{gale_2013} and insulated Physarum 
 wires~\cite{adamatzky_wires}.  
 
 A future Physarum Chip will be a hybrid living-electronic computing device. Being a bio-electronic device the chip 
 will need components to generate waveforms and a clock, the source of regularly spaced pulses, implemented as oscillators.   We experimentally demonstrate that it is possible to implement an electronic oscillator --- 
  a device which converts direct current to alternating current signal --- with living slime \emph{P. polycephalum}.  Experimental setup is outlined in Sect.~\ref{methods}.  Section~\ref{results} presents experimental results on Physarum resistance oscillations, input to output potential transfer function and current dynamics in Physarum oscillators.  Advantages and limitations of living Physarum oscillators are discussed in Sect.~\ref{discussion}.

\section{Materials and Methods}
\label{methods}

\begin{figure}[!tbp] 
\centering
\includegraphics[width=1\textwidth]{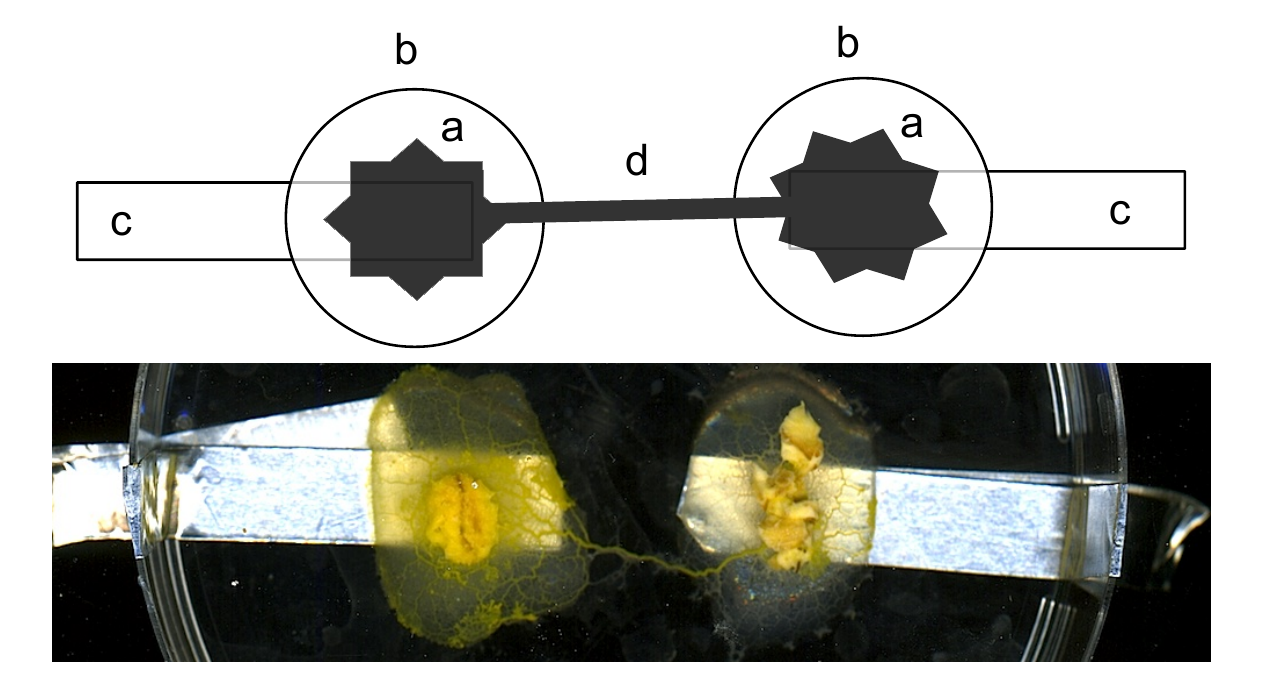}
\caption{Top: A scheme of experimental setup. (a)~Physarum, (b)~agar blobs, (c)~electrodes, (d)~protoplasmic tube. All parts of Physarum 
shown in dark grey form a single cell. Bottom: A snapshot of agar blobs occupied by Physarum and connected by a protoplasmic tube.}
\label{scheme}
\end{figure}

 A scheme of experimental setup is shown in Fig.~\ref{scheme}. Two blobs of agar 
2 ml each (Fig.~\ref{scheme}b) were placed on electrodes (Fig.~\ref{scheme}c) stuck  to a bottom of a plastic Petri dish (9~cm). Distance between proximal sites of electrodes is 10~mm in all experiments. Physarum was inoculated on one agar blob. We waited till Physarum colonised the first blob, where it was inoculated, and propagated towards and colonised the second blob. When second blob is colonised, two blobs of agar, both colonised by Physarum (Fig.~\ref{scheme}a), became connected by a single protoplasmic tube  (Fig.~\ref{scheme}d). In each experiment a resistance, potential and current were measured during 10~min with four wires using Fluke 8846A precision voltmeter, test current 
1$\pm$0.0013 $\mu$A. Direct current potential was applied using Gw Instek GPS-1850D laboratory DC power supply.
Dynamics of resistance of the blobs connected by a single protoplasmic tube was measured in 22 experiments; dynamics of electrical potential difference between blobs and current for direct current potential applied in a range of 2~V to 15~V was measured in 6 experiments for each value of potential difference applied. 

\section{Results}
\label{results}

\begin{figure}[!tbp] 
\centering
\includegraphics[width=1\textwidth]{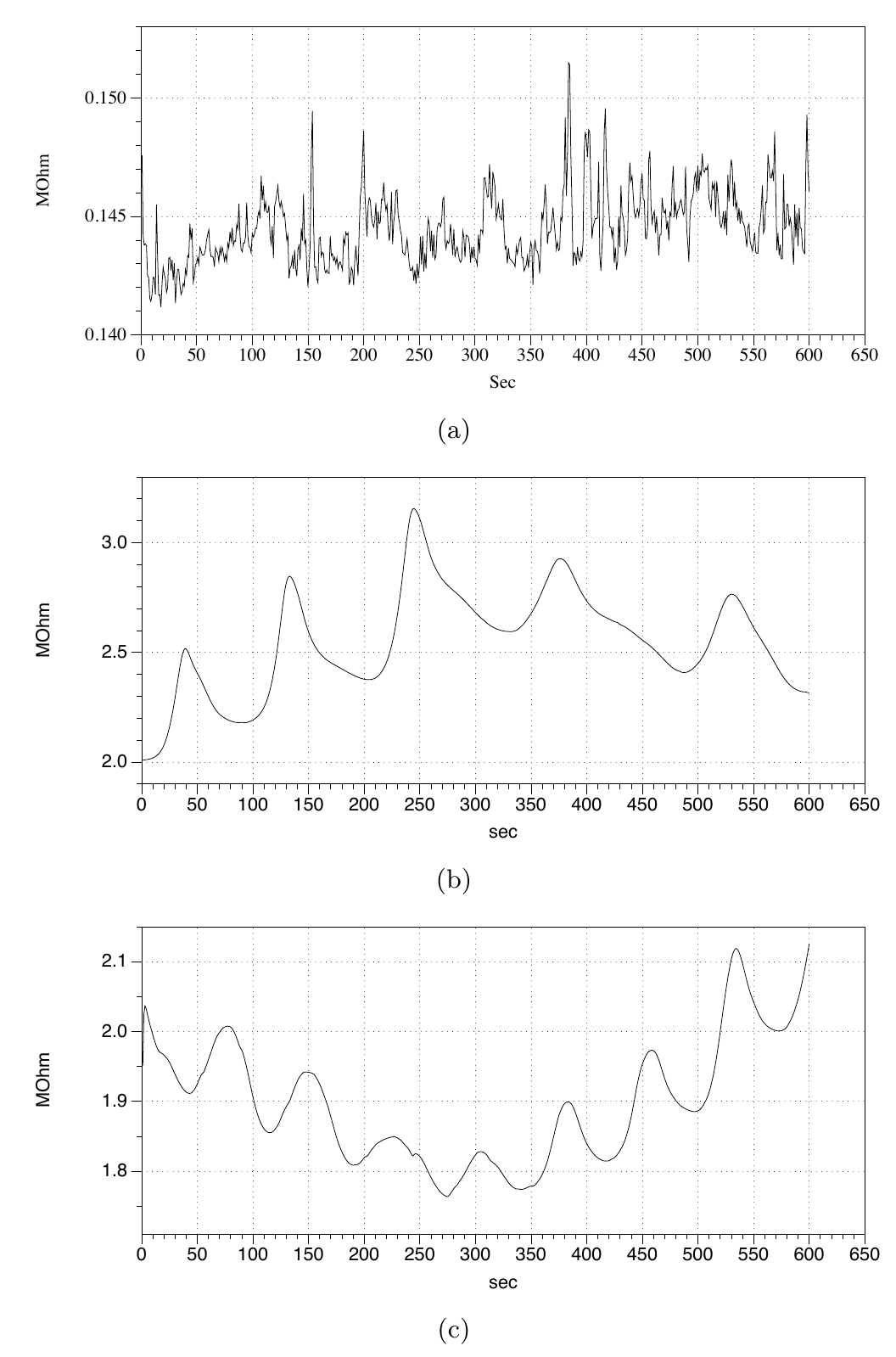}
\caption{Resistance of (a)~of agar blobs, colonised by Physarum, blobs are connected with each other
by a silver wire,  (bc)~agar blobs, occupied by Physarum, blobs are connected with each other by a
protoplasmic tube. The resistance is recorded during 10~min. Vertical axis is resistance in MOhm, and horizontal axis is time in sec.}
\label{exampleresistance}
\end{figure}

\begin{figure}[!tbp] 
\centering
\includegraphics[width=1\textwidth]{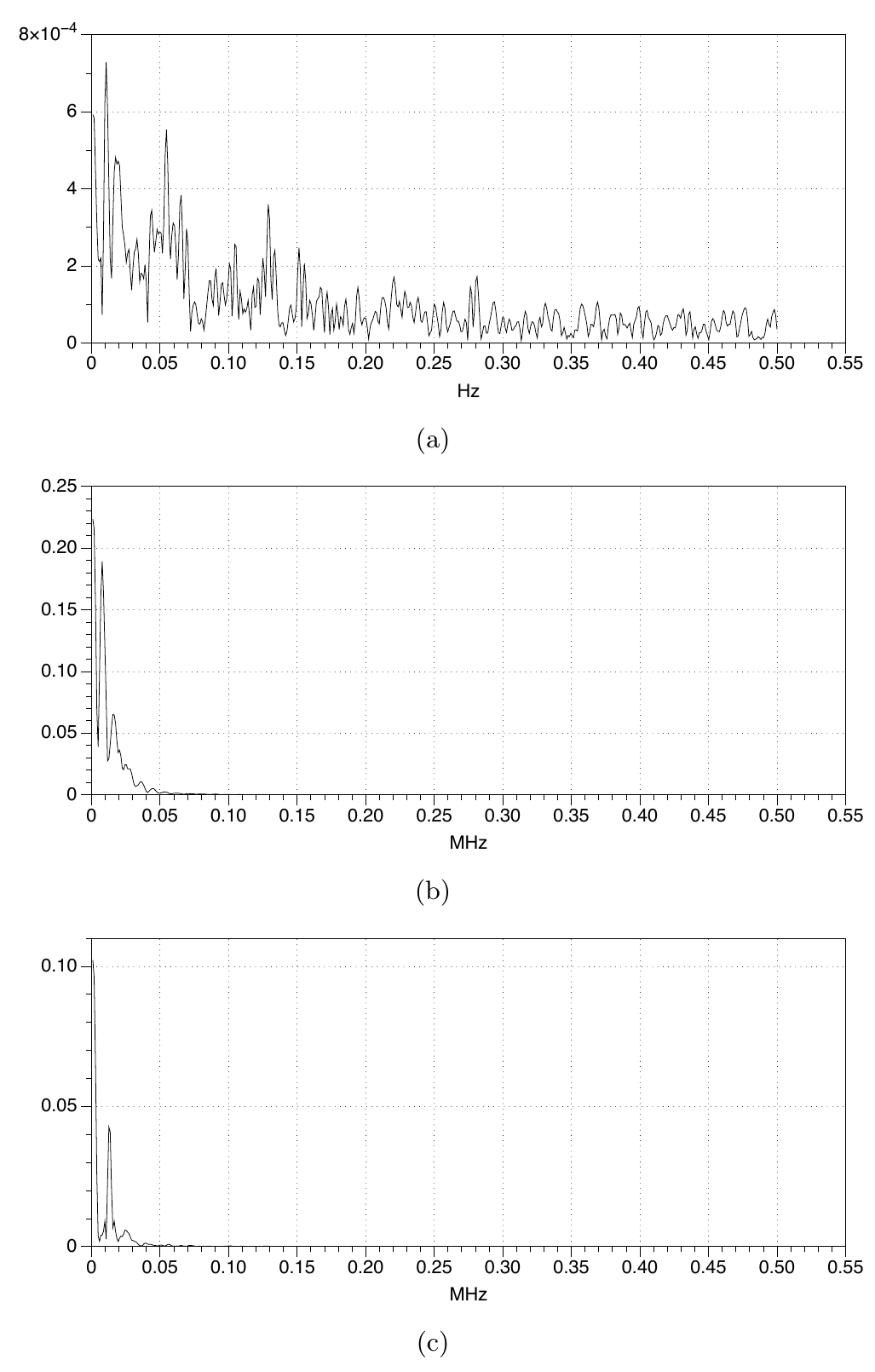}
\caption{Spectra of resistance dynamics  of (a)~of agar blobs, colonised by Physarum, blobs are connected with each other by a silver wire,  (bc)~agar blobs, occupied by Physarum, blobs are connected with each other by a
protoplasmic tube. The spectra are calculated on recordings of resistance shown in Fig.~\ref{exampleresistance}.}
\label{FFTsamples}
\end{figure}

Resistance of two agar blobs colonised by Physarum and connected with each other by a silver wire exhibits quasi-chaotic oscillations (Fig.~\ref{exampleresistance}a) with a wide range of dominating frequencies 
(Fig.~\ref{FFTsamples}a).  When a Physarum propagates from one agar blob to another blob it connects two blobs with 
 a single protoplasmic tube (Fig.~\ref{scheme}b). Average resistance of a 10~mm protoplasmic tube is 3~MOhm, standard deviation 0.715~MOhm.  Resistance of the protoplasmic tube exhibits oscillatory behaviour  (Fig.~\ref{exampleresistance}bc) with highly pronounced dominating frequency (Fig.~\ref{FFTsamples}bc). Average dominating frequency is 0.0137606~Hz,  median dominating frequency is 0.0126953~Hz, standard deviation is 0.0033492429~Hz. 
 The resistance oscillations have average amplitude 0.59~MOhm, standard deviation 0.256~MOhm; minimum amplitude of  resistance oscillations observed was 0.11~MOhm and maximum amplitude 1~MOhm. Oscillation in resistance observed are due to peristaltic contractions of the protoplasmic tube~\cite{sun_2009}, see details in Sect.~\ref{discussion}.

\begin{figure}[!tbp] 
\centering
\includegraphics[width=0.8\textwidth]{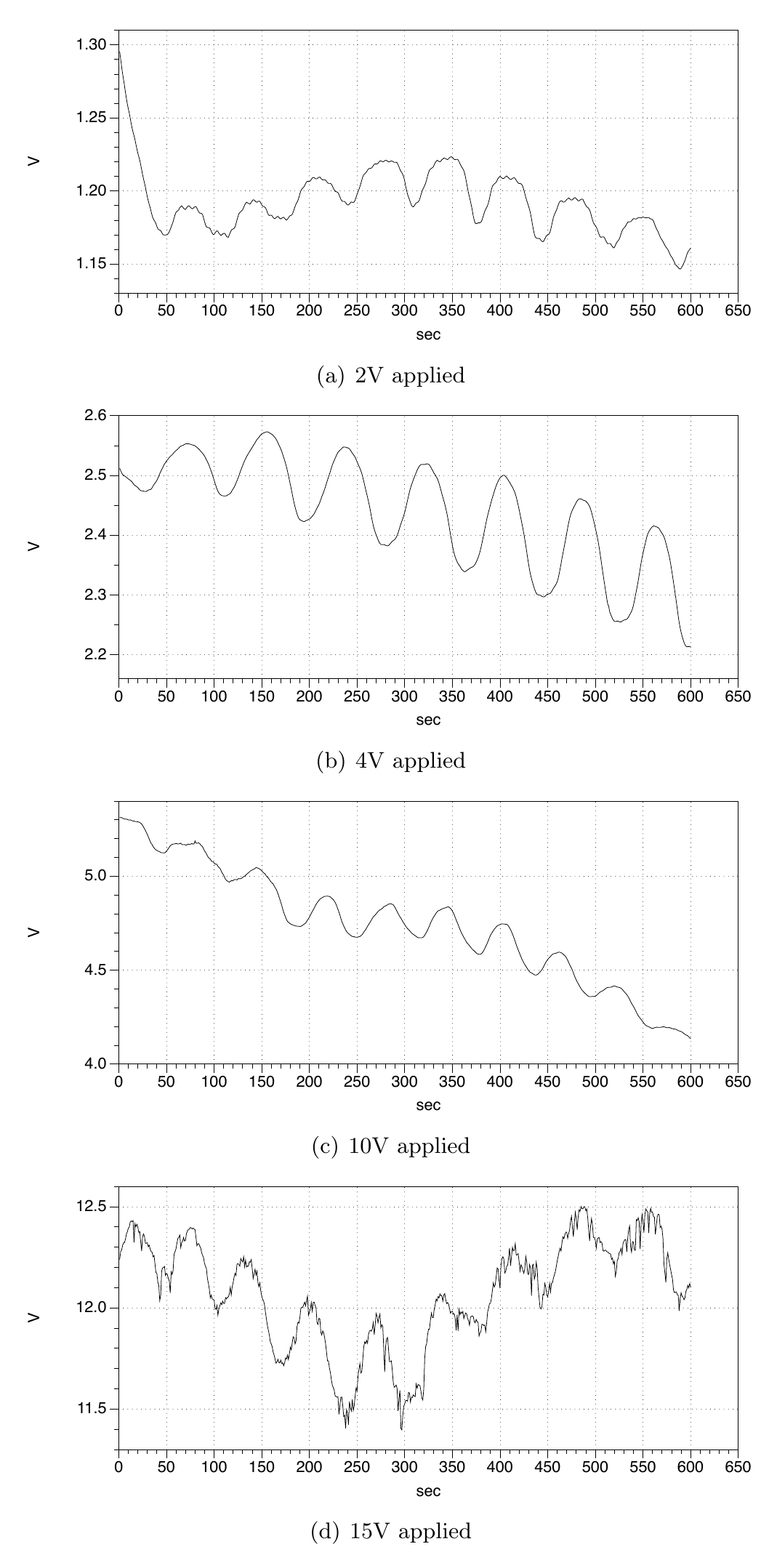}
\caption{Examples of output potential dynamics for various values of applied, input, potential. Vertical axis is potential in V, and horizontal axis is time in sec.}
\label{outputvoltages}
\end{figure}

When we apply a potential to a protoplasmic tube oscillations of the tube's resistance result in oscillation of the output potential as shown in Fig.~\ref{outputvoltages}. Values of average output potential; and frequencies, periods and 
amplitudes of potential oscillations are given in Tab.~\ref{statisticstable}.

\begin{figure}[!tbp] 
\centering
\includegraphics[width=0.8\textwidth]{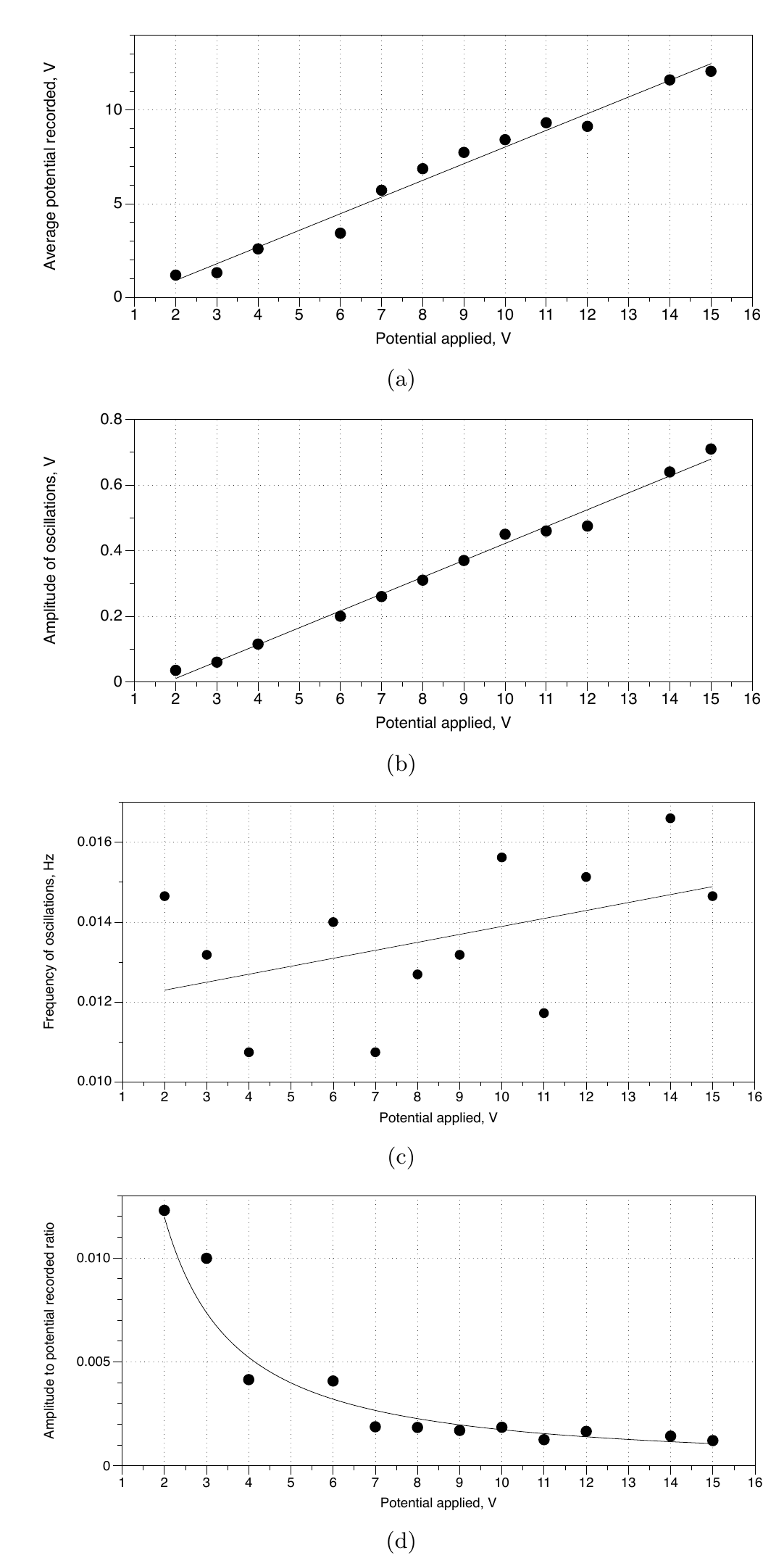}
\caption{Graphs of experimental values of an (a)~average output potential, 
(b)~average amplitude of output potential oscillations, (c)~average frequency of output potential 
oscillations, and (d)~ratio of average amplitude of output potential oscillations to average output potential for 
input potential 2~V to 15~V applied. Trend lines are linear in (abc) and power low in (d).}
\label{relations}
\end{figure}

Average output potential and average amplitude of output potential oscillations grow linearly with increase of 
an input potential (Fig.~\ref{relations}ab). Frequency of oscillations remains almost constant (Fig.~\ref{relations}c), 
Physarum oscillator produces the same frequency oscillations at 2~V and 15~V applied potential.  
A ratio of average amplitude of output potential oscillations to average output potential decreases by a power low with 
increase of input potential (Fig.~\ref{relations}d).

\begin{figure}[!tbp] 
\centering
\includegraphics[width=1\textwidth]{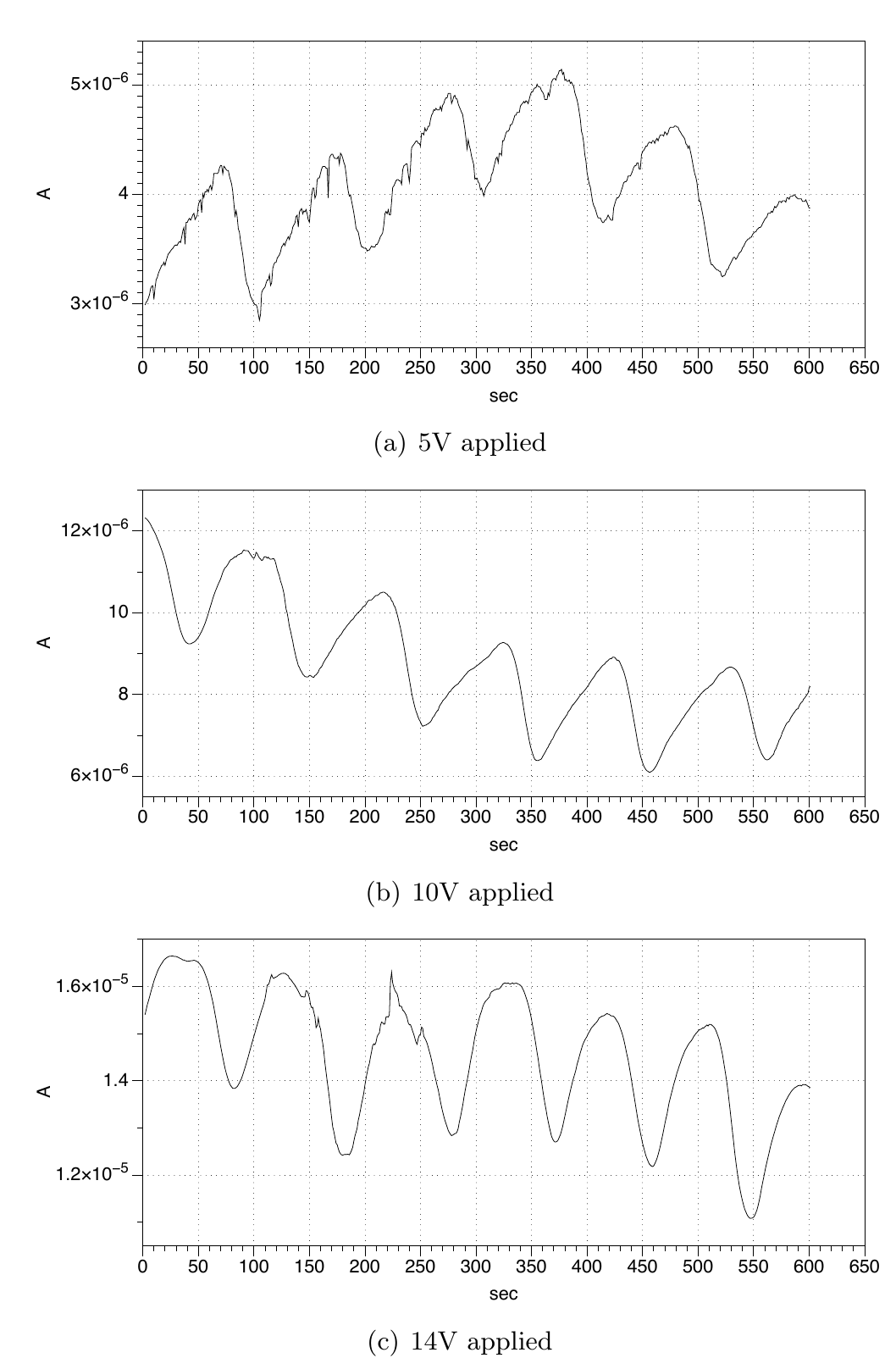}
\caption{Current oscillations in Physarum protoplasmic tubes at (a)~5~V, (b)~10~V and (c)~14~V input potential applied. Vertical axis is current in A, and horizontal axis is time in sec. }
\label{current}
\end{figure}

Examples of oscillations of output current for for 5, 10 and 14~V input potential applied as shown in 
Fig.~\ref{current} and values of current are given in Tab.~\ref{tabcurrent}. 

\section{Discussions}
\label{discussion}

Stability, accuracy, adjustability and ability to produce accurate waveforms are amongst key desirable features of an 
ideal electronic oscillator~\cite{horowitz}.  

\begin{figure}[!tbp]  
\centering 
\includegraphics[width=1\textwidth]{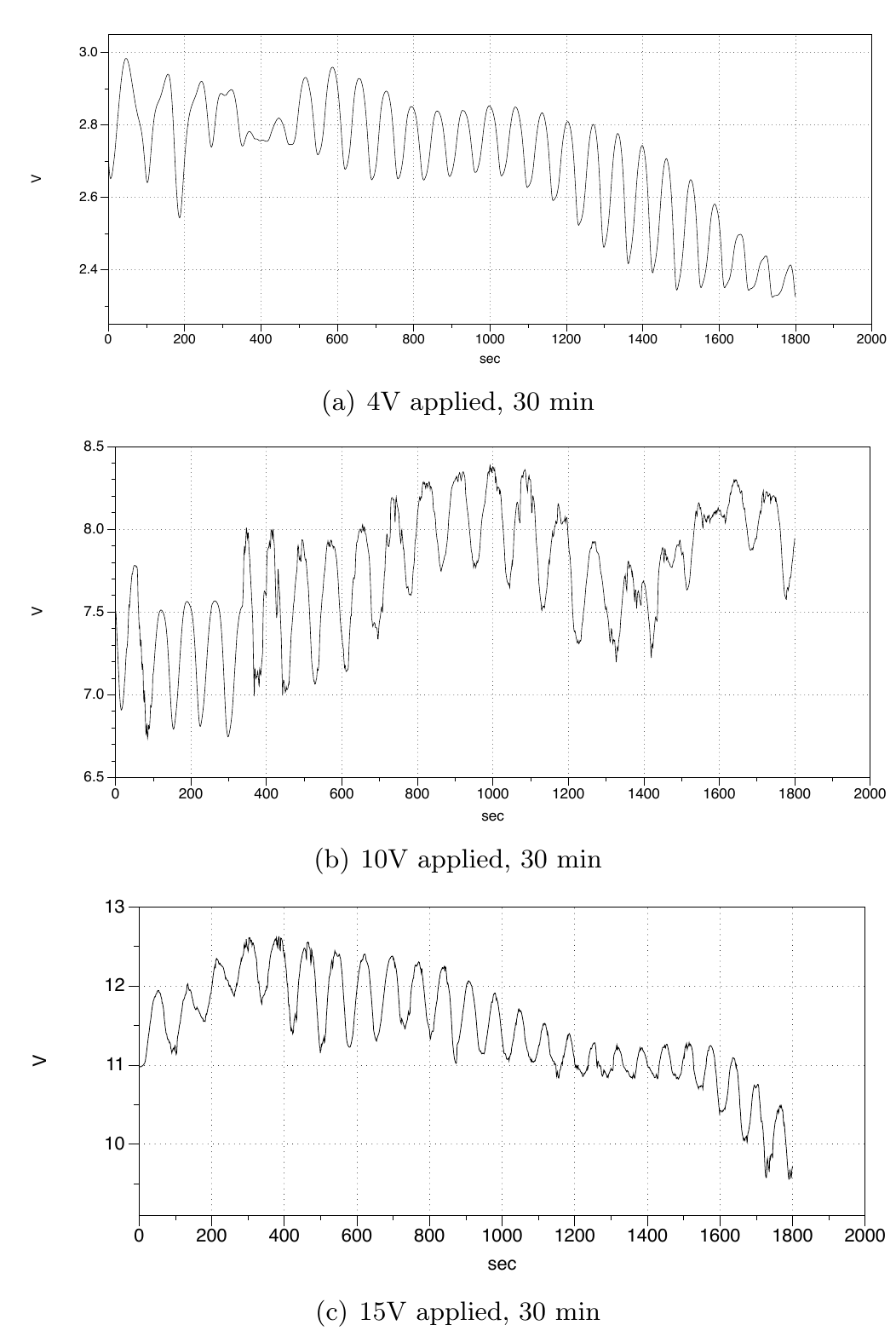}
\caption{Dynamics of output potential of Physarum oscillator during 30~min application of 
input potential of (a)~4~V, (b)~10~V, (c)~5~V. Vertical axis is potential in V, and horizontal axis is time in sec.}
\label{outputvoltageslongrun}
\end{figure}

To test Physarum oscillators' stability we recorded output behaviour of oscillators for 30~min, examples are shown in Fig.~\ref{outputvoltageslongrun}. In such long runs oscillations of an output potential persisted and frequency of oscillations was stable. We observed drifts of the output potential baseline and, sometimes sudden yet short-living, changes in amplitude of oscillations, e.g. between 300th and 500th seconds in example Fig.~\ref{outputvoltageslongrun}a, 1100th and 1500th seconds in example Fig.~\ref{outputvoltageslongrun}b, and 1100th and 1500th seconds in example Fig.~\ref{outputvoltageslongrun}c. Frequency of oscillations is \emph{stable} during over 70\% time  of Physarum oscillator functioning. 

A drift of the electrical potential baseline occurs more likely due to unequal growth of Physarum on agar blobs. Dynamically changing difference in Physarum body mass on reference and recording electrodes leads to a
corresponding changes in resistance of the system and subsequent drift of the background potential. This drift is 
often in a range exceeding the electrical potential oscillation amplitude and therefore must be dealt with. Immediate 
solution would be to involve auxiliary components to  independently measure background resistance or potential, calculate and adjust the baseline potential~\cite{kottke_1999}. Further studies will concern how to control growth of 
Physarum mass on the electrodes. This could possibly be done by illuminating blobs with a strong light every time 
increase of mass is detected. Physarum exhibits photo-avoidance and therefore  its grows will be limited in the illuminated areas.

Protoplasmic tubes of the slime mould exhibit periodic propagations of calcium wave along the tubes.  A calcium ion flux through membrane triggers oscillators responsible for dynamic of contractile activity~\cite{meyer_1979,fingerle_1982}.
The calcium waves are reflected in oscillations of external membrane potential of Physarum and periodic reversing of cytoplasmic flow in the tubes~\cite{iwamura_1949, kamiya_1950, kashimoto_1958, meyer_1979}.  Average dominating frequency of resistance oscillation is 0.0137606~Hz, i.e. period of oscillations is c. 73~sec. This is consistent with our previous findings on periods of electrical potential oscillation and reversing of cytoplasmic flow. Thus, average period of electrical potential oscillation recorded in our experiments is 67~sec~\cite{adamatzky_tactilesensor}, 97~sec~\cite{adamatzky_RGBsensor}, 103~sec~\cite{mayne_2013}, and 115~sec~\cite{whitting_2013}; average 95~sec over four above sets of data~\cite{adamatzky_tactilesensor, adamatzky_RGBsensor,mayne_2013, whitting_2013}. Cytoplasmic flow in tubes reverses with period 54~sec~\cite{adamatzky_schubert}.  Further experiments might deal with establishing an 
exact link between oscillations of a protoplasmic tube's resistance and  peristaltic oscillations of the tube~\cite{sun_2009}.

Frequency of electrical potential oscillations of Physarum can be modified by tactile~\cite{adamatzky_tactilesensor}
and chemical~\cite{whitting_2013} stimuli, by coloured illumination of a protoplasmic 
tube~\cite{adamatzky_RGBsensor},  and by loading Physarum with functional nano-particles~\cite{whitting_2013}.
The correlation between resistance frequency and frequency of electrical potential oscillations allows to speculate
that frequency of resistance oscillation can be also tuned by chemical, optical and tactile control inputs. Thus, we believe the Physarum electrical oscillator is \emph{adjustable}.

Applicability of Physarum oscillators in conventional electronic circuits is limited to none. Physarum oscillators produce 
a very low frequency waveforms and therefore can only be used in computing devices where speed of information processing is not critical. Thus a potential application domain of Physarum oscillators is in self-growing biological computing devices and hybrid bio-silicon devices, and amorphous bio-inspired robots~\cite{melhuish_2001}. 
Another promising application would be in disposable bio-sensors and bio-circuits: living Physarum oscillator can produce stable waveforms for up to 5-7 days; such time frame is sufficient to make reliable measurements and to perform non-time consuming computations.

\section{Conclusions}

A single protoplasmic tube of an acellular slime mould \emph{P. polycephalum} can be employed as a living electronic oscillator which produces stable, accurate, and, in principle, adjustable waveforms. This is because the tube exhibits periodic peristaltic contractions which lead to oscillations of the tube's resistance.

\begin{table}
\caption{Data on Physarum oscillator.}
\begin{small}
\begin{tabular}{cccccc}
Input potential, V	&	Av. freq., Hz	&	Av. period, sec	& Av. amplitude, V	&	Av. output potential, V\\ \hline
2	&	0.01465	&	68.26	&	0.035	&	1.192	\\
3	&	0.01318	&	75.87	&	0.06	&	1.32		\\
4	&	0.01074	&	93.11	&	0.115	&	2.59		\\
6	&	0.014		&	71.43	&	0.2	&	3.43	\\
7	&	0.01074	&	93.11	&	0.26	&	5.72		\\
8	&	0.01269	&	78.80	&	0.31	&	6.87		\\
9	&	0.01318	&	75.87	&	0.37	&	7.74		\\
10	&	0.01562	&	64.02	&	0.45	&	8.42		\\
11	&	0.01172	&	85.32	&	0.46	&	9.32	\\
12	&	0.01513	&	66.09	&	0.475	&	9.13	\\
14	&	0.0166	&	60.24	&	0.64	&	11.61	\\
15	&	0.01465	&	68.26	&	0.71	&	12.07	\\
\end{tabular}
\end{small}
\label{statisticstable}
\end{table}

\begin{table}
\caption{Values of current in  $\mu$A obtained in experiments.}
\begin{tabular}{ccccc}
Potential applied, V & Av. current  & 	St. dev. 	& Av. current amplitude & St. dev \\ \hline
5  			& 4.06 		& 0.53	&  1.23	&  0.29	\\
10 			& 9.04 		& 1.5		&  2.22	&  0.22	\\
14 			& 11.2		& 0.72		&  3.49	&  0.24	\\
\end{tabular}
\label{tabcurrent}
\end{table}

\end{document}